\documentclass[prl,twocolumn,nopacs,nokeys]{revtex4}
\usepackage{amsfonts}
\usepackage{amsmath}
\usepackage{amssymb}
\usepackage{bbm}
\usepackage[utf8]{inputenc}
\usepackage{tensor}
\usepackage{slashed}
\usepackage[centertableaux ]{ytableau}
\usepackage{hyperref}
\usepackage{natbib}
\usepackage{enumerate}
\usepackage{braket,mleftright}
\usepackage{graphicx}
\usepackage[caption=false]{subfig}
\usepackage{subfig}
\usepackage{cleveref}

\begin{document}

\title{Complete analytical solution to the quantum Yukawa potential }
 \author{ M. Napsuciale$^{(1)}$, S. Rodr\'{\i}guez$^{(2)}$ }
\address{$^{(1)}$Departamento de F\'{i}sica, Universidad de Guanajuato, Lomas del Campestre 103, Fraccionamiento
Lomas del Campestre, Le\'on, Guanajuato, M\'exico, 37150.}
\address{$^{(2)}$Facultad de Ciencias F\'isico-Matem\'aticas,
  Universidad Aut\'onoma de Coahuila, Edificio A, Unidad
  Camporredondo, 25000, Saltillo, Coahuila, M\'exico.} 

\begin{abstract}
We present a complete analytical solution to the quantum problem of a particle in the Yukawa potential, using supersymmetry and a 
systematic expansion of the corresponding super-potentials. Results for the critical screening of the ground state improve in several figures 
existing results based on both numerical solutions and approximation methods. Our calculation to order $(a_{0}/D)^{2}$ for the squared 
ground state wavefunction at the origin, which enter in darkonium transitions, yields a correction of $\pi^{4}/216$ to results 
based on variational techniques.
\end{abstract}
\maketitle

The Yukawa potential, given by 
\begin{equation}
V(r)= - \alpha \frac{e^{- r/D}}{r}. 
\label{YP}
\end{equation}
was proposed in Ref.\cite{Yukawa:1935xg} by H. Yukawa as an effective non-relativistic description of the 
strong interactions between nucleons. It appears in many areas in physics and chemistry like atomic physics, plasma physics , 
electrolytes, colloids, and solid state physics \cite{PhysRevA.27.418}  \cite{Debye:1923sr}\cite{RevModPhys.31.569} \cite{PhysRev.125.1131}, 
\cite{PhysRev.134.A1235},\cite{PhysRevB.19.3167}. It is known as Debye-Huckel potential in plasma physics, Thomas-Fermi 
potential in solid state physics or generically as screened Coulomb potential. 
  
The quantum Yukawa potential has a long history, in spite of which, to the best of our knowledge, there is no complete analytical  
solution, either in closed form or as a perturbative expansion. It is well known that for a finite screening there is a finite 
number of bound states \cite{PhysRev.134.A1235} \cite{PhysRev.139.B1428}. The corresponding energy levels depend on the value 
of the screening distance $D$, and 
approximate calculations for some of them are available in the literature, based on variational methods at different 
sophistification level \cite{PhysRev.125.1131} \cite{PhysRevA.9.52}\cite{PhysRevA.8.1138}\cite{PhysRevA.48.220}, perturbation theory 
using the Coulomb potential   \cite{PhysRev.134.A1235} 
\cite{PhysRevA.13.532}\cite{PhysRevA.50.228} \cite{Edwards:2017ndv} \cite{PhysRevA.33.1433} or closely 
related potentials like the Hult\'{e}n potential \cite{PhysRevA.4.1875} \cite{Dutt_1985} as the unperturbed system, or the so called 
logarithmic perturbation theory \cite{Eletsky:1981fm} \cite{Vainberg:1981}  and other methods \cite{PhysRevA.26.1759}\cite{PhysRevA.23.455}
\cite{PhysRevLett.66.1310}\cite{PhysRevA.21.1100}\cite{Moreno:1983nc} \cite{Gonul:2006}\cite{Patil_1984}. 

More recently, the Yukawa potential regained interest as a possibility to solve the core-cusp problem of dark matter 
density profiles  \cite{Loeb:2010gj}\cite{Chan:2013yza}\cite{Khrapak:2003kjw}.  Also, the formation of darkonium is possible for some 
gauge theories of dark matter \cite{MarchRussell:2008tu}\cite{An:2016gad} \cite{Cirelli:2016rnw}\cite{Petraki:2016cnz}\cite{Krovi:2018fdr}.
Our own interest in this problem arose in the study of a $U(1)_D$ gauge theory for tensor dark matter \cite{Hernandez-Arellano:2018sen}\cite{Hernandez-Arellano:2019qgd}. The corresponding phenomenology requires to deeply understand the Yukawa potential and to 
calculate the bound state wave function and their derivatives at the origin. 
  
The intractability of the Yukawa potential and its importance in different fields of physics triggered the numerical studies of this problem
\cite{PhysRevA.1.1577}\cite{PhysRev.159.41}\cite{Li:2006chj}, which shows that Coulomb degeneracy is broken and for 
a given $n$, states with higher $l$ have a higher energy than lower $l$ states. At some point, there is a crossing of levels i.e., states 
of a given $n,l$ have a higher energy than states with $n+1,l^{\prime}$. The critical screening values (those for which a given state 
goes to the continuous) have been also estimated numerically solving the Yukawa potential for $n=0$ to $n=9$ \cite{PhysRevA.1.1577}.
  
In this work, we present a complete analytical solution to the quantum Yukawa problem. The solution is based on the hidden supersymmetry 
 of this potential and on a perturbative expansion of the superpotentials. 

The radial Schrodinger equation for the Yukawa potential 
\begin{equation}
\left[-\frac{\hbar^{2}}{2\mu} \left(\frac{1}{r^{2}}\frac{d}{dr}(r^{2}\frac{d}{dr}) - \frac{l(l+1)}{r^{2}} \right)+V(r) \right]R(r)=E~R(r),
\end{equation}
can be reduced to
\begin{equation}
H_{l}u_{l}=\left[-\frac{d^{2}}{dx^{2}} +v_{l}(x) \right]u_{l}(x)=\epsilon_{l}u_{l},
\label{eom}
\end{equation}
with the effective potential
\begin{equation}
v_{l}(x)= \frac{l(l+1)}{x^{2}} -\frac{2}{x}e^{-\delta x},
\end{equation}
where $x=r/a_{0}$, $R(r)=u(x)/x$ and $\delta = a_{0}/D$, with the Bohr radius, $a_{0}=\frac{\hbar}{\mu c\alpha}$. 
The energy levels  are given by
\begin{equation}
E_{l}=\frac{1}{2}\mu c^{2}\alpha^{2} \epsilon_{l}, 
\end{equation}
where $\mu$ is the reduced mass of the system. We factorize the Yukawa Hamiltonian as 
\begin{equation}
H_{l} = a_{l} a^{\dagger}_{l} + C(l, \delta),
\end{equation}
where
\begin{equation}
a_{l}=-\frac{d}{dx}+W_{l}(x), \qquad a^{\dagger}_{l}=\frac{d}{dx}+W_{l}(x).
\end{equation}
The superpotential $W_{l}$ must satisfy
\begin{equation}
W^{2}_{l}(x,\delta)- W^{\prime}_{l}(x,\delta) + C(l,\delta)= \frac{l(l+1)}{x^{2}} -\frac{2}{x}e^{-\delta x} ,
\label{master0}
\end{equation}
where $ W^{\prime}_{l}\equiv \frac{dW_{l}}{dx}$.
If we succeed in solving the Ricatti equation (\ref{master0}) we also generate a factorization for the partner Hamiltonian defined as
\begin{equation}
\tilde{H}_{l}=a^{\dagger}_{l} a_{l}+ C(l,\delta)=-\frac{d^{2}}{dx^{2}} +\tilde{v}_{l}(x),
\end{equation}
where
\begin{equation}
\tilde{v}_{l}(x)=  W^{2}_{l}(x)+ W^{\prime}_{l}(x) +C(l,\delta) = v_{l}(x) + 2 W^{\prime}_{l}(x).
\end{equation}
The two-component Hamiltonian
\begin{equation}
H=\begin{pmatrix}a^{\dagger}_{l} a_{l}&0 \\ 0&a_{l} a^{\dagger}_{l} \end{pmatrix} ,
\end{equation}
can be written in terms of the charges
\begin{align}
Q_{1}= \begin{pmatrix} 0& -i a_{l} \\ i a^{\dagger}_{l} & 0 \end{pmatrix}, \quad
Q_{2}= \begin{pmatrix} 0&  a_{l} \\ a^{\dagger}_{l} & 0 \end{pmatrix} .
\end{align}
These operators satisfy the $N=2$ supersymmetry algebra \cite{Witten:1981nf}\cite{COOPER1983262}
\begin{equation}
\{Q_{i},Q_{j}\} = 2 \delta_{ij} H, \qquad [Q_{i}, H]=0.
\end{equation}
Explicitly, the Hamiltonian is given by
\begin{equation}
H=\begin{pmatrix} -\frac{d^{2}}{dx^{2}} + U_{+}(x,l) & 0 \\0&  -\frac{d^{2}}{dx^{2}} + U_{-}(x,l)  \end{pmatrix} ,
\end{equation}
with the associated potentials
\begin{equation}
U_{\pm}(x,l)=W^{2}_{l}(x) \pm W^{\prime}_{l}(x).
\end{equation}

Expanding the effective potential in powers of $\delta$ we get 
\begin{align}
W^{2}_{l}(x,\delta)- W^{\prime}_{l}(x,\delta) &+ C(l,\delta)= \frac{l(l+1)}{x^{2}} - \frac{2}{x} \nonumber \\
&+ 2 \delta  - \delta^{2} x + \frac{1}{3}\delta^{3} x^{2} + ... .
\label{master}
\end{align}
The expansion on the right hand side (r.h.s.) of this equation is also an expansion in powers of $x$. The $\delta$-independent 
terms correspond to the Coulomb potential. The ${\cal O}(\delta^{k})$ term on the r.h.s. is ${\cal O}(x^{k-1})$.  Working to 
${\cal O}(\delta^{k})$, we find polynomial solution in $x$ for $W_{l}(x,\delta)$, with the advantage 
that powers of $\delta$ and $x$ are correlated. The general solution can be written as
\begin{align}
W_{l}(x,\delta)&=w_{c}(x,l) + a_{1}\delta+ (a_{2}\delta^{2}+a_{3}\delta^{3}+a_{4}\delta^{4}...) x \nonumber \\
&+ (b_{3}\delta^{3} + b_{4}\delta^{4}+b_{5}\delta^{5}+...) x^{2}  \nonumber \\
&+ ( c_{4}\delta^{4} + c_{5}\delta^{5}+c_{6}\delta^{6} +...) x^{3} +..., \\
C(l,\delta)&=c(l)+y_{1}(l) \delta + y_{2}(l) \delta^{2} +  y_{3}(l) \delta^{3} + ... \label{Lexp}
\end{align}
Here, $w_{c}(x,l) $ is the $\delta$-independent part which corresponds to the Coulomb problem. The coefficients required 
to a given order in $\delta$, can be fixed matching powers of $x$ on both sides of this equation.

We find that to ${\cal O}(\delta^{2})$, the Yukawa problem is factorizable in the sense of Ref. \cite{Infeld:1951mw}. A family of 
supersymmetric Hamiltonians $\{H^{0}(l)\equiv H_{l},H^{1}_{l},H^{2}_{l}..,H^{r}_{l} \}$  with "shape invariant" potentials as described 
in  \cite{Gendenshtein:1984vs} can be built and the spectrum can be straightforwardly  obtained as
\begin{align}
\epsilon_{r,l}&=-\frac{1}{(l+r+1)^{2}} +2\delta \nonumber \\
& - [(l+1)(l+\frac{3}{2}) +3r(r+2(l+1))]\delta^{2},
\end{align}
which when written in terms of the principal quantum number $n=l+r+1$ reads
\begin{equation}
\epsilon_{n,l}=-\frac{1}{n^{2}}+2\delta - \frac{1}{2}[3n^{2}-l(l+1)]\delta^{2}.
\label{E2nl}
\end{equation}
The angular momentum quantum number takes the values $l=n-1, n-2, ..., 1,0$.
The eigenstate $u_{n,n-1}(x)$ satisfies $a^{\dagger}_{n-1}u_{n,n-1}=0$, a condition that can be used to obtain its explicit form as
  \begin{align}
  u_{n,n-1}(x,\delta)&=N_{n,n-1} e^{-\int W_{n-1}(x,\delta)dx} \nonumber \\
  &=N_{n,n-1} x^{n}e^{-\frac{x}{n} +\frac{1}{4}n\delta^{2}x^{2}}.
  \end{align}
States with lower values of $l$ can be obtained iteratively with the aid of the operator $a_{l}$
 \begin{equation}
  u_{n,n-s}(x)=N_{n,n-s} a_{n-s}u_{n,n-s+1},
  \end{equation}
 where $s=2,...,n$, and $N_{n,n-s}$ are $\delta$-dependent normalization factors.

 To  ${\cal O}(\delta^{3})$ and beyond we loose shape invariance. However, supersymmetry is always present and can be used to solve the problem. 
 First, we expect the condition $a^{\dagger}_{n-1} u_{n,n-1}=0$ to be satisfied, which yields
  \begin{align}
  u_{n,n-1}&(x,\delta)=N_{n,n-1} e^{-\int W_{n-1}(x,\delta)dx}\nonumber \\
  &=N_{n,n-1} x^{n} e^{-\frac{x}{n}}  e^{[\frac{n}{2}\delta^{2}-\frac{n}{6}(n+1)\delta^{3}]\frac{x^{2}}{2} - \frac{n}{6} \delta^{3}  \frac{x^{3}}{3}  }. 
  \end{align}
Using this function in Eq. (\ref{eom}) we can check that it is indeed an eigenfunction with eigenvalue
\begin{align}
\epsilon_{n,n-1}&=-\frac{1}{n^{2}}+2\delta-n(n+\frac{1}{2})\delta^{2} \nonumber \\
&+ \frac{1}{3}n^{2}(n+1)(n+\frac{1}{2})\delta^{3}.
\end{align}
Since $H_{l}$ and $H_{l-1}$ are not longer supersymmetry partners, states with lower $l$ cannot be obtained simply applying the
lowering operator $a_{l}$. In order to surmount this difficulty, we solve the supersymmeytric partner 
\begin{equation}
\tilde{H}_{l}\equiv \tilde{H}^{(1)}_{l}\equiv a^{\dagger}_{l}a_{l}+C(l,\delta)=-\frac{d^{2}}{dx^{2}}+\tilde{v}^{(1)}_{l}(x),
\end{equation}
following the same procedure used to solve $H_{l}$ for $l=n-1$. First we re-factorize $\tilde{H}^{(1)}_{l}$ as 
\begin{equation}
\tilde{H}^{(1)}_{l}=\tilde{a}_{l}^{(1)} (\tilde{a}_{l}^{(1)})^{\dagger}+\tilde{C}^{(1)}(l,\delta),
\end{equation}
where
\begin{align}
\tilde{a}_{l}^{(1)}=-\frac{d}{dx}+\tilde{W}_{l}^{(1)}(x), \quad (\tilde{a}_{l}^{(1)})^{\dagger}=\frac{d}{dx}+\tilde{W}^{(1)}_{l}(x).
\end{align}
The new superpotential $\tilde{W}^{(1)}_{l}(x)$ must satisfy an equation similar to Eq.(\ref{master}), but with $\tilde{v}^{(1)}_{l}$ 
on the right hand side. Solving this equation we obtain the solution of this potential for $l=n-2$ as
\begin{align}
\tilde{u}^{(1)}_{n,n-2}(x)&= e^{\int \tilde{W}^{(1)}_{n-2}(x) dx} =x^{n}e^{-\frac{x}{n}} \nonumber \\
&\times e^{\frac{1}{2}n\delta^{2}-\frac{1}{12}n (n^{2}+3 n - 2)\delta^{3} x^{2}- \frac{1}{18}n\delta^{3} x^{3}}.
\end{align} 
The corresponding energy is
\begin{align}
\tilde{\epsilon}^{(1)}_{n,n-2}&=-\frac{1}{n^{2} }+ 2\delta -(n-\frac{1}{2})(n+2)\delta^{2}  \nonumber \\
&+\frac{1}{3}(n-\frac{1}{2})n^{2}(n+5)\delta^{3}=\tilde{C}^{(1)}(n-2,\delta).
\end{align}
 Now we can find the eigenstate of $H_{l}$ for $l=n-2$ using supersymmetry and the double factorization 
  \begin{align}
 \tilde{H}^{(1)}_{n-2}&=\tilde{a}^{(1)}_{n-2}(\tilde{a}^{(1)}_{n-2})^{\dagger}+\tilde{C}^{(1)}(n-2,\delta), \nonumber \\
 &=a^{\dagger}_{n-2} a_{n-2}+C(n-2,\delta).
 \end{align}
 The state $\tilde{u}^{(1)}_{n,n-2}(x)$ satisfy
 \begin{align}
 [a^{\dagger}_{n-2} a_{n-2}+C(n-2,\delta)]\tilde{u}^{(1)}_{n,n-2}=\tilde{\epsilon}^{(1)}_{n,n-2}\tilde{u}^{(1)}_{n,n-2}.
 \end{align}
 Acting on the last equation with $a_{n-2}$ we realize that 
\begin{equation}
u_{n,n-2}=N_{n,n-2} ~a_{n-2}\tilde{u}^{(1)}_{n,n-2}
\end{equation}
is an eigenstate of $H_{l}$ with eigenvalue 
 \begin{equation}
\epsilon_{n,n-2}=\tilde{\epsilon}^{(1)}_{n,n-2}.
 \end{equation}
 Eigenstates and eigenvalues for $l=n-3$ can be calculated applying now this procedure to the superpartner of $\tilde{H}^{(1)}_{l}$.  
 Continuing this process we will eventually reach the lowest $l=0$ level, completely solving the Yukawa problem to order ${\cal O}(\delta^{3}) $. 
 The complete set of eigenvalues to ${\cal O}(\delta^{3}) $ is given by 
 \begin{align}
  \epsilon_{n,l}(\delta) &=  - \frac{1}{n^{2}} + 2\delta - \frac{1}{2} [3n^{2}-l(l +1)]\delta^{2} \nonumber \\
  &+\frac{n^{2}}{6}(5n^{2}+1-3l(l+1))\delta^{3}.
  \label{E3}
  \end{align}
  
The algorithm used to order $\delta^{3}$ can be applied to any order of the expansion of the Yukawa potential. 
We find that the energy levels depend in general of $n^{2}$ and $L^{2}\equiv l(l+1)$ and to order $\delta^{k}$
can be written as
\begin{equation}
\epsilon_{n,l}(\delta)=\sum^{k}_{i=0} \varepsilon_{i}(n^{2},L^{2})\delta^{i}.
\label{Enl}
\end{equation}
The coefficients $\varepsilon_{i}(n^{2},L^{2})$ for $i=0,1,2,3$ are given in Eq. (\ref{E3}) . The next four coefficients in the series are
\begin{eqnarray}
\varepsilon_{4}(a,b)&=&-\frac{a}{96}(77a^{2}+55a-30ab-15b^{2}-6b), \nonumber \\
\varepsilon_{5}(a,b)&=&\frac{a^{2}}{160}(171a^{2}+245a-70ab-45b^{2}-50b+4), \nonumber \\
\varepsilon_{6}(a,b)&=&-\frac{a^{2}}{2880}( 4763 a^3-2070 a^2 b+11580 a^2-945 a b^2 \nonumber \\
&&-2940 a b+1057 a-340 b^3-205 b^2-30 b), \nonumber \\
\varepsilon_{7}(a,b)&=&\frac{a^{3}}{8064}(22763 a^3-10857 a^2 b+84700 a^2 \nonumber \\
&&-4095 a b^2-26145 a b +19677 a-2163 b^3  \nonumber \\
&&- 3843 b^2-2058 b+36).
\label{Coeffk}
\end{eqnarray}

Our formalism yields the numerical factors in the coefficients of the Taylor series given in Eq. (\ref{Coeffk}), but not the $i$-dependence of 
these coefficients, which would allow us to estimate the convergence radius of the series from the limit of the ratio 
$\varepsilon_{i}/\varepsilon_{i-1}$ when $i\to \infty$. However, even in the case of divergent series, we can use the information contained
in the partial sum to a given order $k$,  to reconstruct the complete function $\epsilon_{n,l}(\delta )$. 
Indeed, the appearence of divergent Taylor series is an old problem in quantum mechanics and quantum field theory 
\cite{PhysRev.184.1231} \cite{Bender:1973rz}\cite{ZinnJustin:1980uk}\cite{PhysRevD.23.2916} \cite{Okopinska:1987hp}, and several methods 
are available to sum them up, i.e., to reconstruct the function whose Taylor expansion yields the series \cite{Arteca:1990xe}. These methods 
apply to convergent or divergent series  and the reconstruction is more precise as we take more terms in the Taylor series.
We choose to work with the Pad\'{e} approximants method  \cite{osti_4454325}, which 
is by now a standard technique to analytically continue Taylor series beyond their convergence radii. 
For the series in Eq.(\ref{Enl}), to a given order $k=M+N$ we can always find a rational function
\begin{equation}
[M/N](\delta)=\frac{P_{M}(\delta)}{Q_{N}(\delta)},
\end{equation} 
where $P_{M}(\delta)$ and $Q_{N}(\delta)$ are polynomials of order $M$ and $N$ respectively, such that its Taylor expansion coincides
with the Taylor expansion of $\epsilon_{nl}(\delta)$ to order $k=M+N$. The coefficients of these polynomials are fixed by the coefficients 
$\varepsilon_{i}(n^{2},L^{2})$ in Eq.(\ref{Coeffk}). 

\begin{figure}[h]
\begin{center}
\includegraphics[scale=0.55]{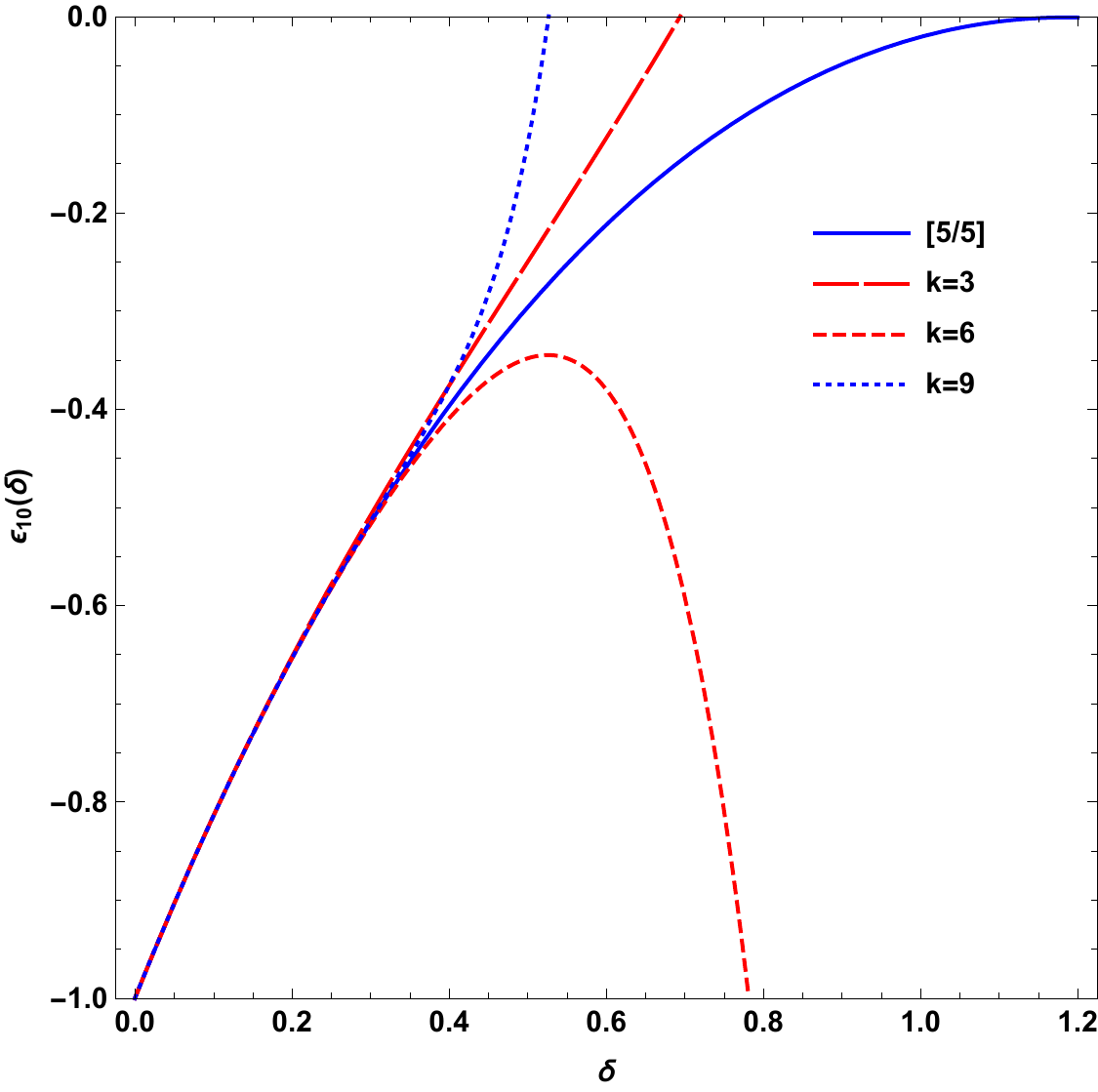}
\end{center}
\caption{Ground state energy of the Yukawa potential calculated with the Taylor series to order $\delta^{k}$ with $k=3,6,9$ and the reconstruction 
of the function $\epsilon_{10}(\delta )$ with the Pad\'{e} approximant $[5/5](\delta)$.}
\label{E10}
\end{figure}

In Fig.\ref{E10} we show the results for the ground state energy as calculated with the Taylor series in Eq.(\ref{Enl}) to order 
$k=3,6,9$ as well as the reconstruction of the function $\epsilon_{10}(\delta)$ with the Pad\'e approximant $[5/5](\delta )$. We use the 
Pad\'{e} approximants built in the {\it Mathematica} package for the calculations in this paper.

The actual value of $\epsilon_{10}(\delta)$ is bounded from above and below by the values 
of the $[(N+1)/N]$ and $[N/N]$ approximants \cite{Arteca:1990xe}, and this fact can be used to estimate the precision in the calculation of 
the energy levels for a given $\delta$. The required precision dictates the order $k=2N+1$ at which is necessary to calculate the Taylor 
series in order to construct the $[(N+1)/N]$ and  $[N/N]$ approximants.  For $N=10$ the energy levels $\epsilon_{nl}$ have an uncertainty of the order of $10^{-7}$ near the critical screening and higher precision in the small $\delta$ region. 
Results based on the numerical solution in \cite{PhysRevA.1.1577} are improved in several figures at this stage. 

One of the most important physical parameters of the Yukawa potential for practical applications is the ground state
critical screening $\delta_{10}$, the value of $\delta$ for which $\epsilon_{10}(\delta)=0$. The numerical solution to the Yukawa problem 
yields $\delta_{10}=1.1906$ \cite{PhysRevA.1.1577}, a value in the large $\delta$ region. 

\begin{figure}[h]
\begin{center}
\includegraphics[scale=0.6]{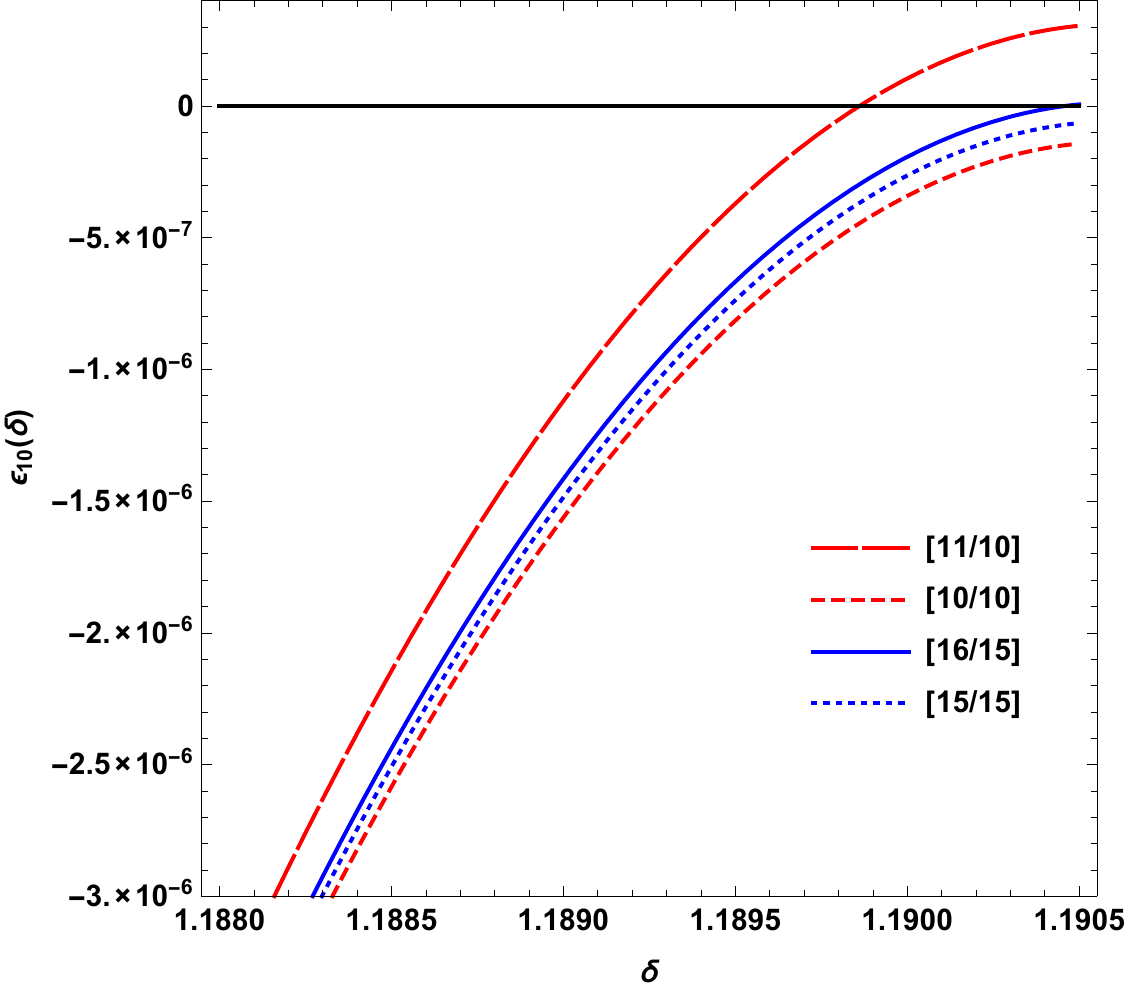}
\end{center}
\caption{Ground state energy of the Yukawa potential near the critical value of $\delta$, reconstructed with the Pad\'{e} approximants $[(N+1)/N](\delta)$ and $[N/N](\delta)$ with $N=10$ and $N=15$.}
\label{cs10}
\end{figure}

In order to calculate this parameter we need to reconstruct  $\epsilon_{10}(\delta)$ in the whole rank of physical values of $\delta$ (those 
for which bound states exist). First, we check numerically that the $[(N+1)/N]$ and $[N/N]$ approximants converge, i.e. that for all values 
of $\delta$, the difference $[(N+1)/N](\delta )-[N/N](\delta )$ reduces as we increase $N$. In Fig. \ref{cs10} we plot the  
$[(N+1)/N](\delta )$ and $[N/N](\delta )$ approximants of the ground state energy in the region near the critical value for $N=10$ and $N=15$, 
which shows that this is indeed the case. 
Then we find numerically the values $\delta^{(N+1)}$ and $\delta^{(N)}$ for which $[(N+1)/N](\delta^{(N+1)})=0$ and  $[N/N](\delta^{(N)})=0$.
These values coincide up to a given figure, which yields the value of $\delta_{10}$. The uncertainty in the calculation is given by the difference
$\delta^{(N+1)}-\delta^{(N)}$. We find that for the $[N/N]$ approximant of the ground state energy, we need to go at least to $N=20$ in order to reach 
the continuum. Using $N=21$ (which requires a calculation of the Taylor series for the ground state energy to order $k=43$) we obtain the value
\begin{equation}
\delta_{10}=1.1906124207(2).
\end{equation}
where the last digit is the uncertainty in the calculation.

Another important parameter of the quantum Yukawa potential is the value of the wavefunction at the origin. It appears in the 
phenomenology of darkonium (non-relativistic bound states of dark matter-dark matter interacting through a Yukawa 
potential \cite{MarchRussell:2008tu}\cite{An:2016gad}\cite{Krovi:2018fdr}). In our formalism it can be
confidently calculated since the $r=0$ limit is well behaved. Using the expansion to order $\delta^{3}$ is enough for this purpose. 
For the ground state, the wave function at this order is given by
\begin {equation}
\psi_{10}(\mathbf{r})=\frac{e^{-x+\frac{1}{12}\delta^{2}(3-2\delta) x^{2} 
-\frac{\delta^{3}}{18}x^{3}}}{\sqrt{\pi a^{3}_{0}(1+\frac{3}{2}\delta^{2}-\frac{11}{6}\delta^{3}})} .
\end{equation}
The systematic calculation of the wave function to order $\delta^{3}$ requires to expand this expression to this order. 
Performing the expansion we find that, at the origin, its square has the value 
\begin{equation}
|\psi_{10}(0)|^{2}=\frac{1}{\pi a^{3}_{0}}(1-\frac{3}{2}\delta^{2}+\frac{11}{6}\delta^{3}).
\end{equation} 
Comparing this result with estimates from variational calculations in \cite{PhysRevA.4.1875} \cite{Krovi:2018fdr} we see that variational 
techniques yield the right sign in the corrections to the Coulomb result but, to order $\delta^{2}$, screening effects in this observable are 
underestimated by a factor $\pi^{4}/216$.

Details of the calculations and a thorough study of the phenomenology of the Yukawa potential based on the present solution will 
be published elsewhere.

\bibliographystyle{prsty}
\bibliography{Yukawa}

\begin{thebibliography}{10}

\bibitem{Yukawa:1935xg}
H. Yukawa, Proc. Phys. Math. Soc. Jap. {\bf 17},  48  (1935).

\bibitem{PhysRevA.27.418}
C.~S. Lam and Y.~P. Varshni, Phys. Rev. A {\bf 27},  418  (1983).

\bibitem{Debye:1923sr}
V. Debye and E. Huckel, Physikalische Zeitschrift {\bf 9},  185  (1923).

\bibitem{RevModPhys.31.569}
H. Margenau and M. Lewis, Rev. Mod. Phys. {\bf 31},  569  (1959).

\bibitem{PhysRev.125.1131}
G.~M. Harris, Phys. Rev. {\bf 125},  1131  (1962).

\bibitem{PhysRev.134.A1235}
C.~R. Smith, Phys. Rev. {\bf 134},  A1235  (1964).

\bibitem{PhysRevB.19.3167}
B. Zee, Phys. Rev. B {\bf 19},  3167  (1979).

\bibitem{PhysRev.139.B1428}
H.~M. Schey and J.~L. Schwartz, Phys. Rev. {\bf 139},  B1428  (1965).

\bibitem{PhysRevA.9.52}
K.~M. Roussel and R.~F. O'Connell, Phys. Rev. A {\bf 9},  52  (1974).

\bibitem{PhysRevA.8.1138}
G.~J. Iafrate, Phys. Rev. A {\bf 8},  1138  (1973).

\bibitem{PhysRevA.48.220}
C. Stubbins, Phys. Rev. A {\bf 48},  220  (1993).

\bibitem{PhysRevA.13.532}
J. McEnnan, L. Kissel, and R.~H. Pratt, Phys. Rev. A {\bf 13},  532  (1976).

\bibitem{PhysRevA.50.228}
O.~A. Gomes, H. Chacham, and J.~R. Mohallem, Phys. Rev. A {\bf 50},  228
  (1994).

\bibitem{Edwards:2017ndv}
J.~P. Edwards {\it et~al.}, PTEP {\bf 2017},  083A01  (2017).

\bibitem{PhysRevA.33.1433}
E.~R. Vrscay, Phys. Rev. A {\bf 33},  1433  (1986).

\bibitem{PhysRevA.4.1875}
C.~S. Lam and Y.~P. Varshni, Phys. Rev. A {\bf 4},  1875  (1971).

\bibitem{Dutt_1985}
R. Dutt, K. Chowdhury, and Y.~P. Varshni, Journal of Physics A: Mathematical
  and General {\bf 18},  1379  (1985).

\bibitem{Eletsky:1981fm}
V. Eletsky, V. Popov, and V. Weinberg, Phys. Lett. A {\bf 84},  235  (1981).

\bibitem{Vainberg:1981}
V. Vainberg, V. Eletskii, and V. Popov, Sov. Phys. JETP {\bf 54},  833  (1981).

\bibitem{PhysRevA.26.1759}
A.~E.~S. Green, Phys. Rev. A {\bf 26},  1759  (1982).

\bibitem{PhysRevA.23.455}
C.~S. Lai, Phys. Rev. A {\bf 23},  455  (1981).

\bibitem{PhysRevLett.66.1310}
S.~L. Garavelli and F.~A. Oliveira, Phys. Rev. Lett. {\bf 66},  1310  (1991).

\bibitem{PhysRevA.21.1100}
C.~S. Lai and B. Suen, Phys. Rev. A {\bf 21},  1100  (1980).

\bibitem{Moreno:1983nc}
G. Moreno and A. Zepeda, J. Phys. B {\bf 17},  21  (1984).

\bibitem{Gonul:2006}
B. Gonul, K. Koksal, and E. Bakir, Physica Scripta {\bf 73},  279  (2006).

\bibitem{Patil_1984}
S.~H. Patil, Journal of Physics A: Mathematical and General {\bf 17},  575
  (1984).

\bibitem{Loeb:2010gj}
A. Loeb and N. Weiner, Phys. Rev. Lett. {\bf 106},  171302  (2011).

\bibitem{Chan:2013yza}
M.~H. Chan, Astrophys. J. Lett. {\bf 769},  L2  (2013).

\bibitem{Khrapak:2003kjw}
S. Khrapak, A. Ivlev, G. Morfill, and S. Zhdanov, Phys. Rev. Lett. {\bf 90},
  225002  (2003).

\bibitem{MarchRussell:2008tu}
J.~D. March-Russell and S.~M. West, Phys. Lett. B {\bf 676},  133  (2009).

\bibitem{An:2016gad}
H. An, M.~B. Wise, and Y. Zhang, Phys. Rev. D {\bf 93},  115020  (2016).

\bibitem{Cirelli:2016rnw}
M. Cirelli {\it et~al.}, JCAP {\bf 05},  036  (2017).

\bibitem{Petraki:2016cnz}
K. Petraki, M. Postma, and J. de~Vries, JHEP {\bf 04},  077  (2017).

\bibitem{Krovi:2018fdr}
A. Krovi, I. Low, and Y. Zhang, JHEP {\bf 10},  026  (2018).

\bibitem{Hernandez-Arellano:2018sen}
H. Hern\'{a}ndez-Arellano, M. Napsuciale, and S. Rodr\'{i}guez, Phys. Rev. {\bf
  D98},  015001  (2018).

\bibitem{Hernandez-Arellano:2019qgd}
H. Hern\'andez-Arellano, M. Napsuciale, and S. Rodr\'\i{}guez, JHEP {\bf 08},
  106  (2020).

\bibitem{PhysRevA.1.1577}
F.~J. Rogers, H.~C. Graboske, and D.~J. Harwood, Phys. Rev. A {\bf 1},  1577
  (1970).

\bibitem{PhysRev.159.41}
C.~A. Rouse, Phys. Rev. {\bf 159},  41  (1967).

\bibitem{Li:2006chj}
Y. Li, X. Luo, and H. Kroger, Science in China Series G {\bf 49},  60  (2006).

\bibitem{Witten:1981nf}
E. Witten, Nucl. Phys. B {\bf 185},  513  (1981).

\bibitem{COOPER1983262}
F. Cooper and B. Freedman, Annals of Physics {\bf 146},  262   (1983).

\bibitem{Infeld:1951mw}
L. Infeld and T. Hull, Rev. Mod. Phys. {\bf 23},  21  (1951).

\bibitem{Gendenshtein:1984vs}
L. Gendenshtein, JETP Lett. {\bf 38},  356  (1983).

\bibitem{PhysRev.184.1231}
C.~M. Bender and T.~T. Wu, Phys. Rev. {\bf 184},  1231  (1969).

\bibitem{Bender:1973rz}
C.~M. Bender and T. Wu, Phys. Rev. D {\bf 7},  1620  (1973).

\bibitem{ZinnJustin:1980uk}
J. Zinn-Justin, Phys. Rept. {\bf 70},  109  (1981).

\bibitem{PhysRevD.23.2916}
P.~M. Stevenson, Phys. Rev. D {\bf 23},  2916  (1981).

\bibitem{Okopinska:1987hp}
A. Okopinska, Phys. Rev. D {\bf 35},  1835  (1987).

\bibitem{Arteca:1990xe}
G. Arteca, F. Fernandez, and E. Castro, {\em {Large order perturbation theory
  and summation methods in quantum mechanics}} (Springer-Verlag, Berlin, 1990).

\bibitem{osti_4454325}
G.~A. Baker, Jr, pp 1-58 of Advances in Theoretical Physics. Vol. I. Brueckner,
  Keith A. (ed.). New York, Academic Press, 1965.  .

\end{thebibliography}

\end{document}